**Operating mode of high pressure straws with high spatial resolution**

K.I. Davkov, V.V. Myalkovskiy, V.D. Peshekhonov, V.D. Cholakov

The article presents results of studying the operating mode of thin-walled drift tubes (straws) at flushing it with a high-pressure gas mixture, which allowed obtaining extremely high spatial resolution for straw detectors. The results of studying the radiation ageing of straws operating in this mode are also described.

**1. Introduction**

Thin-walled drift tubes (TDT or straws) are widely used in accelerator experiments as detecting parts of track detectors at registration of charged relativistic particles, what is mainly determined by their small radiation thickness and good irradiation stability. The examples of such usage are the transition radiation detector – ATLAS Inner Detector tracking system, tracking detectors at COMPASS, LHCb and others [1, 2, 3]. Straw detectors can have large sensitive surface, high gas-tightness and are capable to operate at flushing with a gas mixture at a pressure up to 5 bar.

High pressure straw coordinate detectors are capable to operate both at a proportional or limited proportionality mode and at a self-quenching streamer mode. When registering charged minimum ionizing particles at current detecting mode at the CERN SPS a spatial resolution of near 40 µm was achieved at a gas mixture pressure in the range of 3-4 bar [4, 5]. The article presents the results of investigation of this operation mode, as well as studying straw radiation ageing in the mode.

**2. High pressure straw operating mode**
**2.1. Bench setup**

Straws with an inner diameter of 9.53 mm were wound by two kapton strips. The inner strip was made of carbon-doped XC-160 foil with 40 µm thickness; the outer one was made of HN50 12.5 µm thick foil with ~ 0.2 µm thick aluminium coating. The straw thickness was ~ 60 µm. A gold-plated tungsten wire with a diameter of 30 µm and a resistance of 70 Ω per meter was employed as an anode.

The straw operating mode investigation was carried out at the test bench with irradiation by gamma-ray quanta with the energy of 5.9 KeV from the Fe-55 source and by electrons with the energy of 3.55 MeV from the Ru-106 source. The straws were flushed with a gas mixture $ArCO_2$ (80/20) at its absolute pressure 3 bar.

Amplifiers based on an MSD-2 circuit with a gain of 35 mV/µA and a rise time of 4 ns identical to those used in the radial coordinate measurement [5, 6] were used for the anode signal registration. Pulses from the amplifier output entered the analog-to-digital converter of DRS4 [7], which digitized input signal at a rate of 5 GHz and stored signal amplitude and shape; then they were transferred to the PC for processing amplitude spectra. At registering high energy electrons from the Ru-106 source the event selection was made on the base of matching signals from the straws and a scintillator counter. The set-up scheme is presented in Fig.1

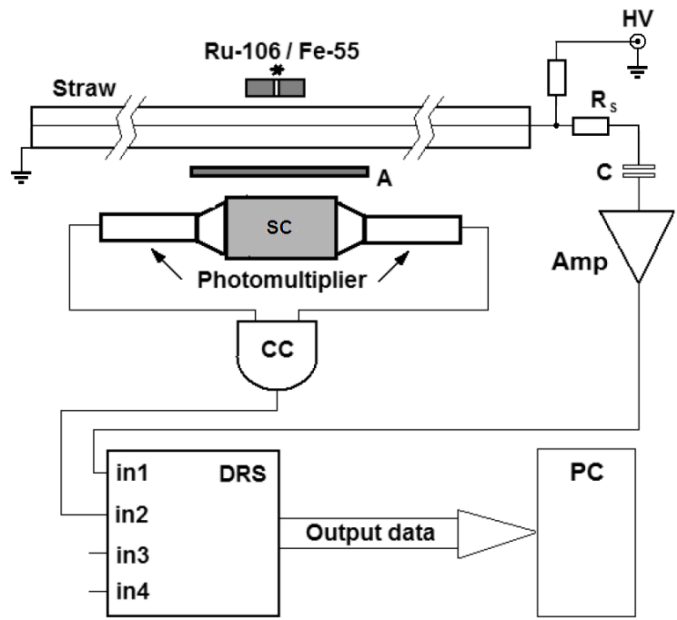

Fig.1. Block diagram of the experiment set-up for studying the operating mode of the straw, where straw is the thin-walled drift tube under study, A – the absorber of low energy electrons, SC – the scintillator counter with two photomultipliers, CC – the coincidence circuit, AMP – the current amplifier, $R_s$ – the series resistor for matching the straw with the amplifier.

**2.2. Investigation of the straw operating mode**

Straw operating modes were examined at a pressure of 3 bar within the anode voltage range of 2.6-3.2 kV, in which the spatial resolution σ was increasing from ~60 to ~40 µm with the anode voltage increase [5].

It has been known that with the increase in gas mixture pressure in drift detectors the possibility of setting a high current mode, capable to convert into a so called self-quenching streamer (SQS) mode, rises [8]. Whereas at a normal pressure, an anode wire diameter of 30 µm and a gas mixture used in the straw SQS mode is almost not observed up to the maximal possible anode voltage.

**2.3. Registration of gamma-ray quanta from the Fe-55 source**

The signal spectra of the Fe-55 source gamma-ray quanta from the straw at a pressure of its gas filling of 3 bar and the anode voltage within the range of 2.6-3.2 kV are shown in Fig.2. At a anode voltage of 2.6 kV (the straw energy resolution is about 33%) and 2.65 kV there can be observed a total-absorption peak (5.9 KeV) and an escape peak (2.95 KeV), which indicates the straw operates in the proportional gain mode. At a voltage ranging from 2.7 to 2.8 kV the escape peak begins to merge with the total-absorption peak, which indicates a transition from the limited proportionality mode to the saturation signal mode. At a voltage of about 2.9 kV high current signals begin to appear and their quantity becomes dominating as the voltage increases up to 3.05 kV, and at a voltage of 3.2 kV the straw almost fully operates in high current mode. The spectra shown in Fig.2 are in good agreement with the spectra from [8].

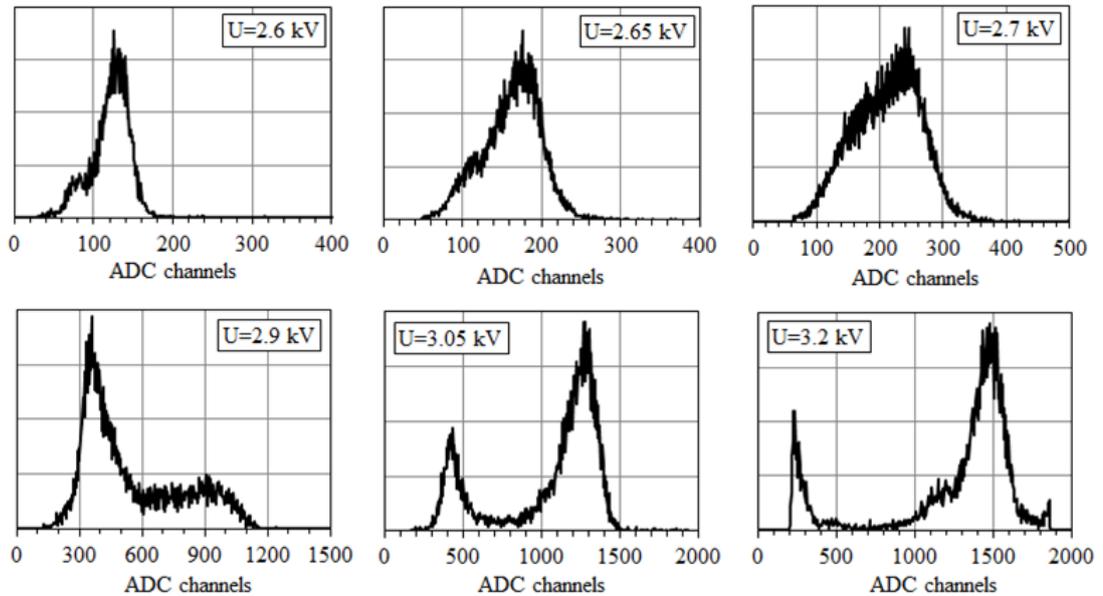

Fig. 2. Fe-55 source signals spectra at different anode voltages. The gas filling pressure is 3 bar.

The dependence of the γ-quantum signal amplitude on the anode voltage is presented in Fig. 3. Lower curve 1 shows the signal values in the proportional mode (at a voltage below ~2.85 kV) and further up to ~3.2 kV in limited proportionality/saturated modes. The transition mode from low current to high current signals is observed in the range of ~2.9 – 3.2 kV, further signals in SQS mode are observed (Curve 2). The signal amplitudes increase with respect to signals in the proportional mode with a factor up to 10.

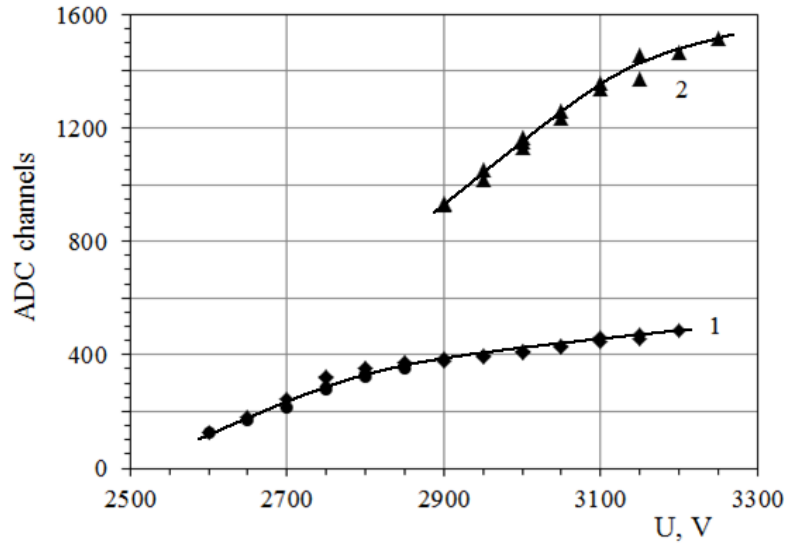

Fig. 3. Signals in different modes of registration of the Fe-55 source gamma-ray quanta with respect to the anode voltage. The gas mixture $ArCO_2$ (80/20) at a pressure of 3 bar.

The quantitative ratio of the low and high current signals (curves 1 and 2, respectively) and the change in the current value in the straw with the anode voltage increase (curve 3) are shown in Fig. 4. Beginning with the anode voltage of ~2.8 kV the high current signals appear in amount of several percent, and at a voltage of ~3.2 kV they become saturated and produce a space charge in the irradiation area, that affects the straw local efficiency. Curve 3 shows the average current in the straw as a function of the anode voltage. It is seen that in the voltage ranging from 2.8 to 3.2 kV the average current increases by a factor of 35 at a fixed gamma-ray quantum flux, rising from ~4 up to ~140 nA. At a voltage of 3.05 kV the current increase is not more than 10 with respect to a voltage of 2.8 kV, which is in agreement with the signal amplitudes correlation in Fig. 3. The ratio of low current signals to high current ones at a voltage of 3.05 kV is 20/80%.

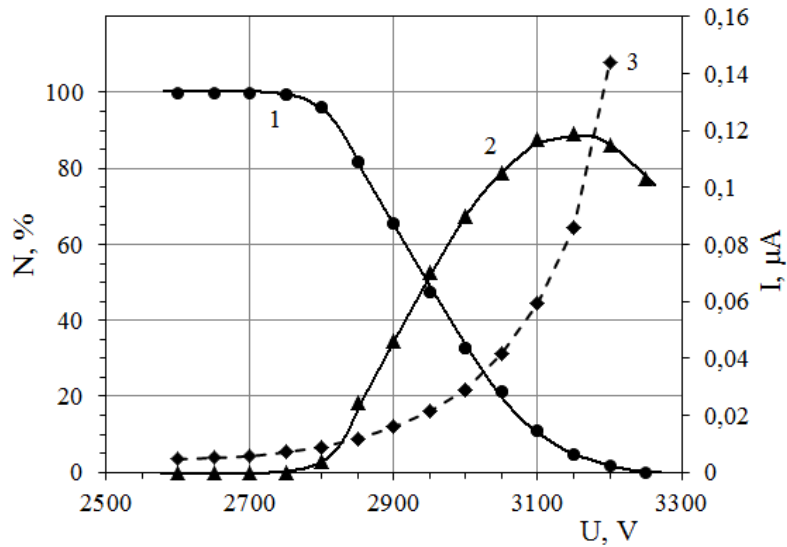

Fig. 4. The signal ratio in different modes of its registration in the straw as a function of the anode voltage. Curve 1 gives the signals from gamma-ray quanta in the proportional or limited proportionality mode. Curve 2 gives the signals in the high current mode. Point graph 3 is the current in the straw as a function of the voltage. The Fe-55 source, the gas mixture $ArCO_2$ (80/20) at a pressure of 3 bar.

### 2.4. Registration of electrons from the Ru-106 source

Dependences analogue to those obtained at the registration of gamma-ray quanta were obtained at the registration of high energy electrons from the Ru-106 source. The dependences of signal amplitudes on the anode voltage at the registration of electrons with the energy of 3.55 MeV are shown in Fig. 5. It can be seen that the transition from low current to high current signals (lower 1 and upper 2 curve, respectively) begins with the anode voltage 100 V higher than that at the registration of gamma-ray quanta.

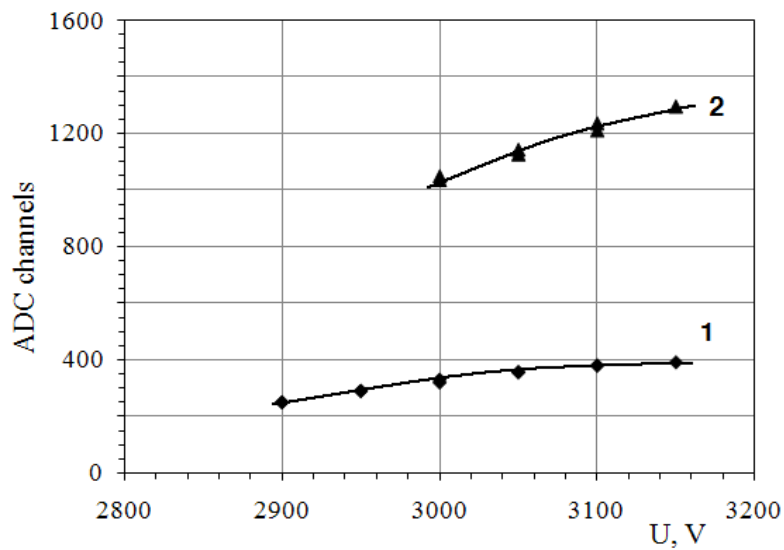

Fig. 5. Signals from the Ru-106 source as a function of the anode voltage. Gas mixture $ArCO_2$ (80/20) at a pressure of 3 bar. Curve 1 gives signals in the proportional or limited proportionality mode. Curve 2 gives signals in the high current mode.

The quantitative ratio of the low and high current signals (curves 1 and 2, respectively) and the change in the current value in the straw with the anode voltage increase (curve 3) at the electron registration are shown in Fig. 6. Beginning with the anode voltage of ~2.9 kV the increase in high current signals is observed; they come up to ~60% at a voltage of 3.15 kV (Fig. 6). At a voltage of 3.05 kV the ratio is 30/70%. The same ratio is observed for the anode voltage of ~2.9 kV at the registration of the Fe-55 source quanta. The point graph in Fig. 6 shows the average current value per one registered event depending on the anode voltage. The value makes 50 nA at a voltage of 3.05 kV, but it is ~2 times skewed upwards as it also comprises the average current from non-detected low-energy electrons with the energy of 39.2 KeV emitted from the source.

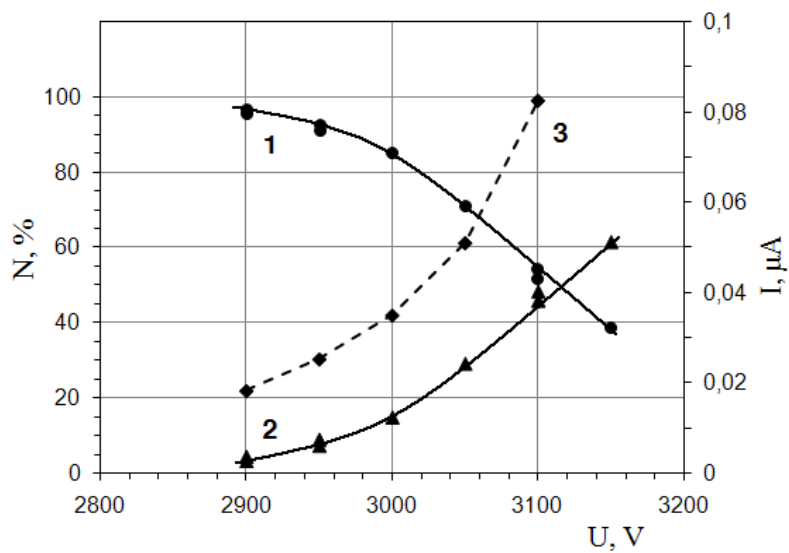

Fig. 6. The signal ratio in different modes of its registration in the straw as a function of the anode voltage. The Ru-106 source, the gas mixture $ArCO_2$ (80/20) at a pressure of 3 bar. Curve 1 gives the signals from high energy electrons in the proportional or limited proportionality mode. Curve 2 gives the signals in the high current mode. Point graph 3 is the current in the straw as a function of the voltage.

**2.5. Summary**

For the gas mixture $ArCO_2$ (80/20) at a pressure ranging from ~3 to 4 bar the straw operation is possible in the mode of transition from the limited proportionality (saturation) mode to the high current mode. With that both types of signals are registered in their stable ratio and the ratio value depends on the anode voltage. The spatial resolution of the straw in this operating mode of the detector can be improved up to ~40 µm [5]. In this case current increase factor in the straw is not more than 10.

It should be mentioned that at registration of gamma-ray quanta and charged particles at similar values of their energy losses switching to the transient mode is observed for quanta at a lower anode voltage.

## 3. Straw ageing study
### 3.1. Bench set-up

An essential factor affecting efficiency of detectors is their radiation ageing. The working medium of straws is renewed constantly, therefore the ageing effects of gas-filled detectors determined mostly by the charge accumulated in them stem from polymerization of sediments on the anode and/or the cathode as a result of possible chemical reactions with active residuals which leads to change in the detector's parameters.

To test the straws' readiness for long-term operation in the transient mode their radiation stability was checked by X-ray irradiation at the test bench.

At the test bench (Fig. 7) there were placed the investigated and monitor prototypes containing the 11.3 cm long straws identical to those described above. The first prototype's straw was irradiated along the whole its length by the γ-ray quantum beam of homogenous intensity from the roentgen tube with a copper anode at a voltage of 9 kV. The other prototype's straw served as a monitor detector (MD) and was used for obtaining comparative amplitude characteristics from the straw under X-ray irradiation when it was tested. The monitor straw with the Fe-55 collimated source permanently situated in its center was irradiated at a load of ~ 100 Hz per 1 cm of the anode length, which eliminated the risk of ageing effects. Both straws were placed close to each other and were flushed sequentially with the gas mixture $ArCO_2$ (80/20) at a rate of ~20 $cm^3$/h and an absolute pressure of 3±0.02 bar, which eliminates possibility of difference in their gas gains due to changes in the partial pressure of the gas mixture's components and external parameters – ambient temperature and pressure.

The straw was irradiated by X-rays at the anode voltage of 3.05 kV, an average current of ~3 µA with an average load of ~ 340 kHz (30 kHz/cm of the anode length).

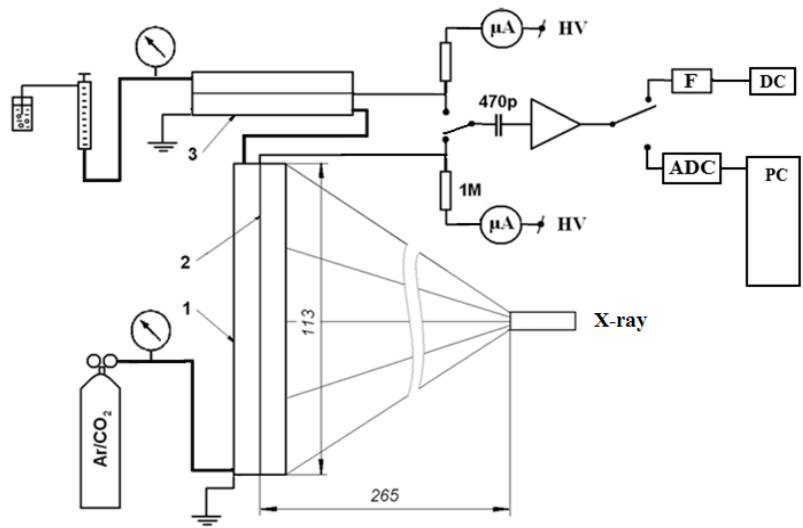

Fig.7. Block diagram of the experiment test bench: 1 –straw under investigation with anode (2); 3 – monitor counter; F – shaper; DC – digital counter; ADC – amplitude-digital converter; PC - computer.

## 3.2. Results

Before the investigation began the test mode for the irradiated straw had been set at the gas mixture pressure of 1 bar and gas gain $\sim 2\times 10^4$. The signal amplitude of the Fe-55 source quanta throughout the length of the straws made 100 mV. Then irradiation uniformity with respect to the first straw length was checked by scanning with quanta from the RT through a slit collimator. The difference in the irradiation intensity in the middle and at the ends of the straw did not exceed 15%.

With every accumulated ~0.5 C/cm of the charge in the irradiated straw it was tested by measuring the signal amplitudes and energy resolution throughout the anode length and these values were compared to those of the monitor straw.

For ~2600 hours of irradiation the average charge per 1 cm of the straw length made 4.2 C. The results of scanning the monitor and irradiated straws are presented in Fig. 8 by curves 1 and 2, respectively. There can be seen a low radiation ageing effect, that consisted in reduction of signal amplitude down to ~8% along the irradiated straw towards the gas mixture flow. Degradation of the energy resolution was not observed.

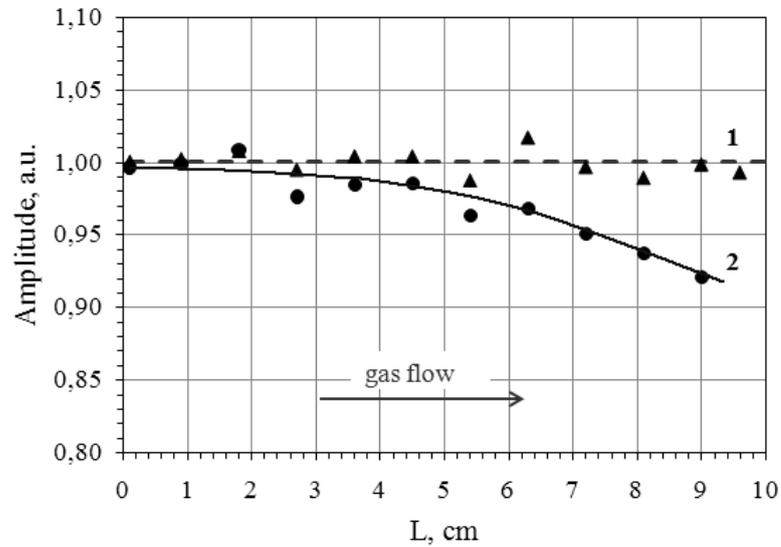

Fig. 8. Signal amplitudes along the length of the monitor straw (curve 1) and the irradiated straw (curve 2). Pressure is 1 bar, gas gain is ~$2 \times 10^4$. Average accumulated charge per 1 cm of the straw length is ~ 4.2 C. The arrow points the gas flow direction.

The gas mixture (volume of ~8.1 cm$^3$) in the irradiated straw at a flow rate of 20 cm$^3$/h was completely changed in 24 minutes; i.e. the irradiation of the gas in the 1 mm thick orthogonal to the anode section had increased ~$10^3$ times by the moment of its release. It shows the possibility of increased polymerization along the straw due to low flow rate when the straw is irradiated throughout its length. Besides it indicates that in this case the transient mode current at the registration of gamma-ray quanta is higher than at the registration of minimum ionizing particles as it was described above.

Thus the results of the irradiation test with gamma-ray quanta with energies of 8 KeV showed feasibility of the high pressure straw for the long-term operation in the transient mode.

**Conclusion**

The study of the transient mode between the low current and high current modes for straws filled with $ArCO_2$ gas mixture at a pressure of 3 bar showed its feasibility for high-precision registration of charged particles. The transient mode does not develop in the self-quenching streamer mode at the pressure within this range and at an anode diameter of 30 µm or less and also has high stability and enough radiation tolerance.

We thank V.I. Davkov for his participation in the radiation ageing study.